\documentclass[aps,prc,twocolumn,reprint,superscriptaddress,nobibnotes,nofootinbib]{revtex4-2}

\usepackage{amsmath,amssymb}
\usepackage{graphicx}
\usepackage{braket}
\usepackage[hidelinks]{hyperref}
\usepackage{xcolor}
\usepackage{placeins}
\usepackage{xspace}

\newcommand{\pA}{p/A}
\newcommand{\trentobase}{T\ensuremath{_\mathrm{R}}ENTo}
\newcommand{\trento}{\trentobase\xspace}
\newcommand{\ftrento}{\trentobase-Sea\xspace}
\newcommand{\ptrento}{\trentobase-Sea\xspace}

\begin{document}
\raggedbottom

\title{Introducing Parton-Level Fluctuations in the \ftrento Model}

\author{Jacob Reinach}
\author{Andreas Kirchner}
\affiliation{Department of Physics, Duke University, Durham, NC 27708, USA}

\date{\today}

\begin{abstract}

A variety of experimental and theoretical data suggest that initial condition models for heavy-ion collisions contain a degree of sub-nucleonic fluctuations. While current models fit the number and size of these fluctuations to data, we use nuclear parton distribution functions to constrain sub-nucleonic structure and fluctuations in \ftrento, an extension of \trento. Sampling the momentum fraction distributions of quarks and gluons in heavy ions, we use perturbative quantum chromodynamics to collide nuclei at the parton level, and we introduce an infrared center-of-momentum damping scale which is tuned to proton-proton cross sections. We find that our model successfully reproduces the $\varepsilon_2\{2\}$ of data-calibrated models except in central collisions, and an analysis of the power spectrum and granularity of the pre-equilibrium fireball shows that the number of low-$x$ partons in a nucleon has a significant effect on the local transverse geometry. Our model is largely insensitive to variations of the infrared parameters, as such changes are absorbed into the calibration of the infrared regulator. We expect that this work will aid future research on soft-hard correlations and constraining different nPDF sets. 
\end{abstract}

\maketitle

\section{Introduction}
\label{sec:intro}

Microseconds after the Big Bang, it is conjectured that the Universe was filled with a nearly perfect fluid of deconfined quarks and gluons, a hot and dense state called the quark-gluon plasma (QGP). While the transition to the QGP is predicted by lattice QCD~\cite{Aoki:2006we,Borsanyi:2010bp,Bazavov:2011nk,Bazavov:2014pvz} and the QGP has been established experimentally~\cite{STAR:2005gfr,PHENIX:2004vcz}, information about its properties and initial conditions is inferred after the freeze-out and hadronization of the QGP, from the subsequent streams of particles which are detected in colliders as indirect evidence. Theorists currently model heavy-ion collisions using multistage simulations whose hydrodynamic evolution is calibrated to real collider data through Bayesian inference~\cite{Bernhard:2019bmu,JETSCAPE:2020mzn,Nijs:2020ors}. The first ingredient in these simulations are the so-called initial conditions, where programs like \trento~\cite{Moreland:2014oya} construct an initial entropy-density or energy-density profile from participating nuclear matter. To date, widely used models that include nucleonic substructure contain a fixed number of constituents with fluctuating weights~\cite{Moreland:2018gsh,Loizides:2016djv,Bozek:2016kpf}. 

The handling of nucleonic substructure has a significant effect on the accuracy of a model. Hydrodynamic evolution converts initial geometry and spatial anisotropy, $\varepsilon_n$, into a momentum anisotropy as measurable flow $v_n$, where $v_n \propto \varepsilon_n$ for $n = 2,3$~\cite{Ollitrault:1992bk,Gardim:2011xv,Niemi:2012aj}. For central collisions and odd harmonics, these initial geometries are dominated by fluctuations rather than by the average geometry~\cite{Alver:2010gr}. This suggests that the granularity of the initial state matters for observables downstream. 

This idea is supported by Bayesian calibrations~\cite{Giacalone:2021clp,Nijs:2022rme} in which it is found that smaller nucleons are favored by the data. In particular, a study by Giacalone \textit{et al.}~\cite{Giacalone:2021clp} found that the nucleon width should be approximately 0.4~fm, unlike earlier calibrations which tended towards larger nucleon widths~\cite{Bernhard:2019bmu,Moreland:2018gsh}. There is also experimental evidence suggesting the need for sub-nucleonic structure in initial state models. The PHENIX collaboration found that constituent-quark-participant scaling described transverse energy distributions better than nucleon-level modeling did, motivating an initial state with partonic degrees of freedom~\cite{Eremin:2003qn,PHENIX:2013ehw,PHENIX:2015tbb}. Recent results from the ALICE collaboration further support the need for sub-nucleonic structure and finer granularity in initial condition models through measurements of flow in light-ion collisions, which seems to favor a sub-nucleon width of approximately 0.1--0.2~fm~\cite{ALICE:2025luc}.

There exist a variety of initial state models such as IP-Glasma, EKRT, and fixed-constituent \trento which include fluctuations and, in most cases, sub-nucleonic structure~\cite{Schenke:2012wb,Schenke:2012fw,Gale:2012rq,Mantysaari:2016ykx,Eskola:1999fc,Niemi:2015qia,Moreland:2018gsh}. IP-Glasma is a pre-equilibrium model built on color glass condensate (CGC) saturation physics. IP-Glasma uses the IP-Sat dipole model to construct colliding gluon fields, which are then evolved using classical Yang--Mills equations. Each nucleon is given a number of gluon ``hot spots'' in this model, with a width of about 0.1~fm constrained by data~\cite{Mantysaari:2016ykx,Mantysaari:2016jaz,Mantysaari:2017cni}. In EKRT models, perturbative QCD (pQCD) minijets are computed using nuclear-modified parton distribution functions (nPDFs). Saturation sets a local scale that fixes the initial energy density of a pre-equilibrium profile, as well as the early formation time. While standard EKRT does not contain nucleonic substructure, the recent Monte Carlo version (MC-EKRT) adds three hot spots of width 0.15--0.2~fm with per-nucleon energy conservation~\cite{Kuha:2024kmq,Hirvonen:2024zne}. This is similar to fixed-constituent \trento models, which also set a fixed number of constituents in nucleons $n_c$, and assign fluctuating weights from a gamma distribution, where constituent size, weight, and number are fit to data~\cite{Moreland:2018gsh}.

To date, each of these models either fixes or fits the number and size of sub-nucleonic sources. In this work, we introduce ``\ftrento,''~\cite{Kirchner:2025yuo} in which the momentum-based weights of constituents are sampled from nPDFs, which also constrains the number of constituents. We sample quarks and gluons from the proper nPDFs for distribution around nucleons in colliding ions. These partons are distributed around nucleons using an $x$-dependent nucleon width motivated by Gribov diffusion~\cite{Frankfurt:2010ea,Strikman:2003gz}, and we set a single infrared center-of-momentum damping scale $s_0$, also referred to as the infrared cutoff scale or the damping scale, which damps low-energy binary parton collisions. We introduce two terms, the $x$-dependent parton placement width $W(x)$ and a power-law infrared regulator, as phenomenological choices, while the momentum fractions and number of partons are constrained by the nPDFs. We compare this model's eccentricity decomposition and power spectrum to previous iterations of \trento, and isolate the role of nPDFs through an analysis of the reduced thickness profiles and the fraction of holes in pre-equilibrium fireballs. The present work is restricted to the two-dimensional initial-state geometry at mid-rapidity and does not include longitudinal structure, pre-equilibrium dynamics, or hydrodynamic evolution. 

In Sec.~\ref{sec:framework}, we outline the physical framework in this model, followed by a description of our sampling techniques in Sec.~\ref{sec:sampling}. In Sec.~\ref{sec:results}, we present our findings, and we discuss future directions and conclusions in Sec.~\ref{sec:conclusions}.

\section{Physical Framework}
\label{sec:framework}
\begin{figure}[!tbp]
    \centering
    \includegraphics[width=1\linewidth]{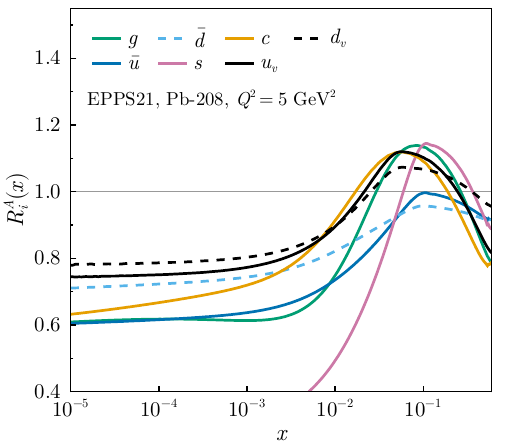}
    \caption{Nuclear modification factor $R_i^A(x)$ for a variety of parton species in Pb-208 from the EPPS21 dataset with $Q^2 = 5$~GeV$^2$. $R_i^A$ is the ratio of the parton density in a bound proton to that in a free proton at a given $x$. At low-$x$, nPDFs sample fewer partons due to shadowing, while there is a peak at higher momentum fractions in nPDFs from anti-shadowing.}
    \label{fig:PDF_vs_nPDF}
\end{figure}

In this work, we sample the species and momentum fractions of a nucleon's constituents from nPDFs, leading to event-by-event fluctuations in the number of constituents per nucleon as the momentum budget of a nucleon is used up. This provides the sub-nucleonic fluctuations favored by data~\cite{Alver:2010gr,Bernhard:2019bmu,Giacalone:2021clp}. The momentum distribution of quarks and gluons within nucleons is quantified here using parton distribution functions (PDFs) from the CT18ANLO set~\cite{Hou:2019efy,Buckley:2014ana}, and from the EPPS21 nPDFs~\cite{Eskola:2021nhw}, which include the nuclear modification factor $R_i^A$, where $A$ is the mass number and $i$ is the flavor of a parton. Each species of parton $i$ has a distribution dependent on $A$ and the momentum transfer scale $Q^2$. Here, we define $Q^2$ as the negative of the square of the four-momentum transfer. These distributions give the number density of partons carrying Bjorken momentum fraction $x$. Mathematically, these distributions are
\begin{equation}
    f_i^{\pA}(x,Q^2) = R_i^A(x,Q^2)\,f_i^{p}(x,Q^2),
\end{equation}
with $f_i^p(x,Q^2)$ referring to the free proton distributions for flavor $i$, $R_i^A(x,Q^2)$ as the nuclear modification factor, and $f^{\pA}_i(x, Q^2)$ referring to the bound proton distributions for flavor $i$ in the nucleus $A$. The nuclear modification factor is depicted for a variety of parton species in Fig.~\ref{fig:PDF_vs_nPDF}. This plot quantifies the ratio of the parton densities in a bound proton to those in a free proton, and quantifies nuclear effects such as shadowing and anti-shadowing, which occur at small $x$ ($x \lesssim 10^{-2}$) and intermediate $x$ ($x \sim 0.1$), respectively. In this work, the shadowing of low-$x$ partons has a significant effect, as seen below.

For initial-state models, every collision is dependent on the kinematic variable $\sqrt{s_{NN}}$, which represents the total center-of-mass energy of a particle collision per nucleon. For any combination of $Q^2$ and $\sqrt{s_{NN}}$ there is a minimum Bjorken momentum fraction $x_\mathrm{min}=Q^2/s_{NN}$ representing the smallest possible value of the Bjorken momentum fraction a parton can carry through deep inelastic scattering processes. For $\sqrt{s_{NN}} = 5$~TeV and $Q^2 = 5$~GeV$^2$ used in this work, $x_\mathrm{min} = 2.0\times10^{-7}$.

When determining the sub-nucleonic structure in nuclei, we first model the valence quarks within every nucleon. For a proton, this is two up quarks and a down quark, and for a neutron, it is two down quarks and one up quark. We sample the momentum fractions of these valence quarks from the proper proton and neutron nPDFs. In low energy collisions, valence quarks carry much larger momentum fractions than other partons. However, at high energies it is rare that the valence quarks carry the entirety of a single nucleon's momentum budget. We then sample the momentum fractions of gluons and quark-antiquark pairs which constitute a virtual sea of partons in each nucleon and carry the remainder of the nucleon's total momentum. The details of this process are further outlined in Sec.~\ref{sec:sampling}.

\begin{figure*}[!tp]
    \centering
    \includegraphics[width=\textwidth]{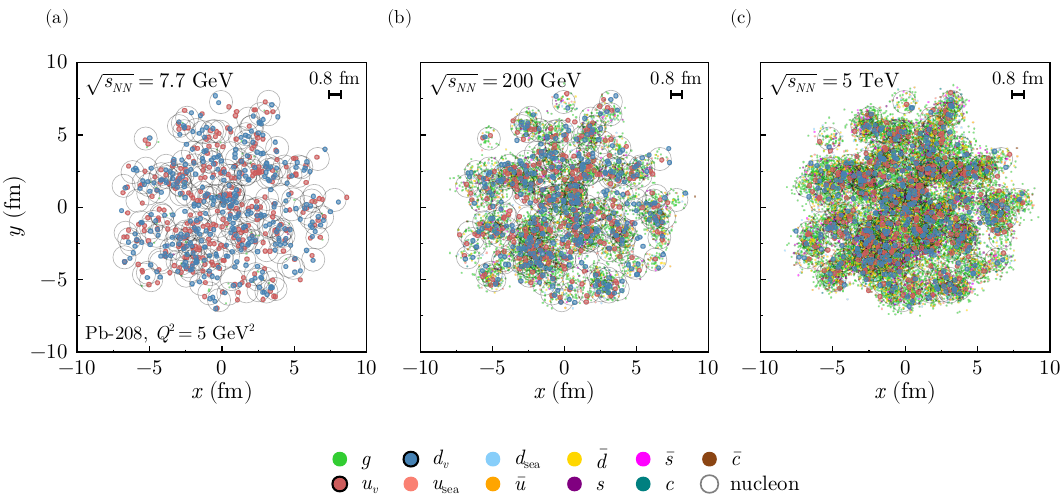}
    \caption{Transverse positions of partons and nucleons in a single Pb-208 nucleus at three different collision energies. The black rings mark each nucleon with a radius of $2w = 0.8$~fm, twice the nucleon width $w = 0.4$~fm, which encloses about 86\% of a Gaussian distribution of width $w$ (a radius of $w$ would enclose only 39\%). Low-$x$ partons, which are placed with a wider $W(x)$, can lie outside the rings. The valence quarks are shown in the more opaque red and blue circles, while virtual sea partons are represented by more transparent colors, with gluons shown in green. Across the same nuclear geometry, lower collision energies have far fewer sea partons, and the collision at $\sqrt{s_{NN}}=7.7$~GeV has nucleons that are almost entirely made up of just the three valence quarks. At $\sqrt{s_{NN}}=5$~TeV, the transverse plane is filled with soft gluons, as well as an increased number of heavier quark species, since $x_\mathrm{min} = Q^2/s_{NN}$ reaches the region where the gluon density rises steeply. This fills in the nucleon profile significantly.}
    \label{fig:transverse}
\end{figure*}

After obtaining the momentum distribution of partons in nucleons, we also need to determine the spatial distribution of both nucleons and partons in a nucleus. We sample the nucleon positions according to a Woods--Saxon distribution, using two separate radii for protons and neutrons. The proton radius is taken as the charge radius, e.g. $R_p = 6.62$~fm~\cite{Moreland:2019szz,DeVries:1987atn} for Pb-208, while the neutron radius is $R_n = R_p + r_\mathrm{skin}$ to account for the neutron skin with e.g. $r_\mathrm{skin} = 0.21$~fm~\cite{Moreland:2019szz} for Pb-208. All nucleons are subject to a minimum distance condition $d_\mathrm{min}$ to prevent unphysical geometric overlap. We then distribute partons around each nucleon center using a Rayleigh distribution about a momentum-dependent nucleon width $W(x)$ which grows as a parton's $x$ decreases. The Rayleigh distribution is defined as
\begin{equation}
    \label{eq:Rayleigh}
    P(\delta)\,d\delta = \frac{\delta}{W^2(x)}\,\exp\!\left(-\frac{\delta^2}{2\,W^2(x)}\right) d\delta,
\end{equation}
with $\delta$ being the amount of radial offset that a parton has from the nucleon center and $P(\delta)d\delta$ is the probability of finding a parton at a radial offset between $\delta$ and $\delta + d\delta$. Each parton is modeled by a Gaussian of width $\nu$ taken to be $\nu = 0.15$~fm such that it can be defined with the thickness function
\begin{equation}
    T_\nu(\vec{r} - \vec{r}_k) = \frac{1}{2\pi\nu^2}
    \exp\!\left( -\frac{|\vec{r} - \vec{r}_k|^2}{2\nu^2} \right),
    \label{eq:Tnu}
\end{equation}
where $\vec{r}$ is a point in the transverse plane and $\vec{r}_k$ is the transverse position of parton $k$.
In Fig.~\ref{fig:transverse}, we show the transverse positions of partons and nucleons in a single Pb-208 nucleus at three different collision energies. The black rings represent nucleons, while the valence quarks and virtual sea partons are represented by more opaque colors. Across the same nuclear geometry, we find that low collision energies have far fewer low-$x$ partons, and the collision at $\sqrt{s_{NN}}=7.7$~GeV has nucleons that are mostly made of valence quarks. However, at $\sqrt{s_{NN}}=5$~TeV the transverse plane is filled with soft gluons (in green), as well as an increased number of heavier quark species. This is expected because larger energy collisions lower the value of $x_\mathrm{min}$ into a region where gluon density rises steeply, causing more low-$x$ gluons to be sampled.

\trento provides the transverse energy density from collisions at mid-rapidity. The nuclei involved are highly Lorentz-contracted, so the eikonal approximation for a Glauber model allows us to consider only the 2D, transverse dynamics of a collision with independent scattering. For two colliding nuclei $A$ and $B$, we refer to each individual parton with the labels $k\in A$ and $l \in B$ and each has flavor $i$ or $j$ respectively. Following Ref.~\cite{Moreland:2019szz}, we define the participant overlap function of interacting partons
\begin{equation}
\begin{split}
    T_{kl}(b_{kl}, \nu) = \int d^2 r \; T_\nu(\vec{r} - \vec{r}_k)\, T_\nu(\vec{r} - \vec{r}_l) \\
    = \frac{1}{4\pi\nu^2} \exp\!\left( -\frac{b_{kl}^2}{4\nu^2} \right),
\end{split}
\end{equation}
where $\vec b_{kl} = \vec r_k - \vec r_l$ and can be considered the partonic impact parameter. The likelihood that two partons collide can then be written as
\begin{equation}
    P_\mathrm{coll}(b_{kl},\nu) = 1 - \exp\!\big(-\sigma_{k_i l_j}(\hat s_{kl})\, T_{kl}(b_{kl}, \nu)\big),
    \label{eq:pcoll}
\end{equation}
where $\sigma_{k_i l_j}(\hat s_{kl})$ represents the effective cross section of colliding partons as a function of the Mandelstam center-of-mass energy squared $\hat s_{kl} = x_k x_l s_{NN}$, where $x_k$ and $x_l$ are the momentum fractions of partons $k$ and $l$.

We determine the effective cross section using perturbative quantum chromodynamics (pQCD) and the integrated $2\to2$ matrix elements with Debye screening in the $t$- and $u$-channels. The Debye mass is set by $\mu_D^2 = s_0/4$, with $s_0$ being the infrared cutoff scale, a tuned quantity introduced in this model as a replacement for low-$x$ cutoffs on individual partons. We introduce a damping factor $\left(\frac{\hat s_{kl}}{\hat s_{kl}+s_0}\right)^{\xi}$ to suppress collisions below the infrared cutoff scale $s_0$, the region in which perturbation theory no longer applies even though there are many parton pairs in this regime. Ultimately, we express the effective parton-level cross sections as
\begin{equation}
\begin{split}
\sigma_{k_i l_j}(\hat s_{kl}) &= \left(\frac{\hat s_{kl}}{\hat s_{kl}+s_0}\right)^{\xi} \\
&\quad\times \frac{\pi\,\alpha_s^2(\hat s_{kl}+s_0)}{\hat s_{kl}^2}\, I_{k_i l_j}\!\left(\hat s_{kl},\ \mu_D^2\right),
\end{split}
\label{eq:sigma}
\end{equation}
where $\xi$ is a phenomenological variable tested over a range of values, $I_{k_i l_j}\!\left(\hat s_{kl},\ \mu_D^2\right)$ are the integrated matrix elements from pQCD~\cite{Combridge:1977dm}, and $\alpha_s(\hat s_{kl}+s_0)$ is the running QCD coupling constant. We define this coupling constant similar to Ref.~\cite{ParticleDataGroup:2024cfk},
\begin{equation}
\alpha_s(\hat s_{kl} + s_0) = \frac{12\pi}{(33 - 2n_f)\,\ln\!\left(\dfrac{\hat s_{kl} + s_0}{\Lambda_\mathrm{QCD}^2}\right)}
\end{equation}
with the characteristic one-loop coupling scale $\Lambda_\mathrm{QCD} = 0.3$~GeV, and the active number of quark flavors $n_f = 4$.

\begin{figure}[!tbp]
    \centering
    \includegraphics[width=1\linewidth]{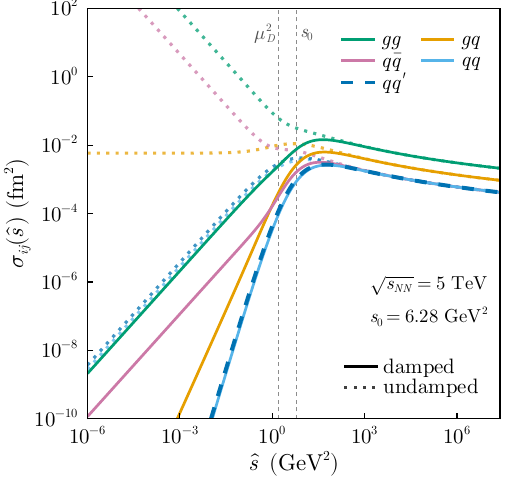}
    \caption{Damped and undamped parton-level cross sections by channel with $s_0 = 6.28$~GeV$^2$ and $\xi = 2$. The cross sections vanish quickly below the infrared cutoff $s_0$ and the corresponding mass screening scale set by $\mu_D^2 = s_0/4 = 1.57$~GeV$^2$. The undamped cross section for gluon-gluon interactions grows at low $\hat s$, saturating the overall collision cross section without any infrared cutoff scale. All channels share the same damping factor in Eq.~\eqref{eq:sigma}, but fall off with different slopes because their matrix elements differ.}
    \label{fig:CrossSections}
\end{figure}

In order to determine the infrared cutoff scale $s_0$, we tune its value such that the inelastic proton-proton cross section given in this model replicates measured proton-proton cross sections~\cite{Loizides:2017ack,dEnterria:2020dwq}.\footnote{The tuned value of $s_0$ depends on $\sqrt{s_{NN}}$ and on the model parameters ($w$, $\nu$, $\xi$, $\mu_D^2$, $x_0$), so it must be recalibrated whenever any of these change.} Using the value of $s_0$ tuned to $pp$ collisions, we reproduce the Pb+Pb inelastic cross section that matches the measured cross section found by experiments with ALICE~\cite{ALICE:2022xir}. For this work, we choose $\xi = 2$, the smallest exponent for which the cutoff stays physical and the observables stop depending on $\xi$. Further details of this process are outlined in Sec.~\ref{sec:sampling}. The resulting damped and undamped parton-level cross sections are shown by channel in Fig.~\ref{fig:CrossSections} with $s_0 = 6.28$~GeV$^2$.
The cross sections in Fig.~\ref{fig:CrossSections} vanish quickly below the infrared cutoff $s_0$ and the corresponding mass screening scale set by $\mu_D^2$, and the rate at which the cross sections vanish is set by the parameter $\xi$. For undamped cross sections, particularly gluon-gluon interactions, the effective cross section grows at low momenta. This presents a nonphysical behavior driven by the failure of pQCD to describe this energy range. Though each channel is damped by the same proportion in Eq.~\eqref{eq:sigma}, the cross sections fall off with different slopes due to the different matrix elements of each channel.

Given the parton-level cross sections and participant overlap functions, we collide partons probabilistically using Eq.~\eqref{eq:pcoll}. Every parton involved in a collision is labeled a participant, and deposits its energy proportional to its momentum fraction $x_k$ at its position in the transverse plane $\vec r_k$. The overall energy density for a given collision is given by summing over all participant partons. Following Refs.~\cite{Moreland:2014oya,Kirchner:2025yuo}, the fluctuating thickness field of each nucleus is
\begin{equation}
\tilde{T}_{A(B)}(\vec{r}) = \sum_{k \,\in\, \mathrm{part}_{A(B)}} x_k \, T_\nu(\vec{r} - \vec{r}_k),
\end{equation}
with $k\in\mathrm{part}_{A(B)}$ referring to participant partons. This expression does not carry the usual gamma fluctuations in previous \trento models, and each contributing parton is scaled by its own momentum fraction instead. In turn, this model utilizes the inherent fluctuations of parton numbers and momentum fractions when sampled from nPDFs, providing a physics-motivated source of fluctuations without the need to calibrate a gamma distribution. 

This model follows the reasoning of the original \trento~\cite{Moreland:2014oya} hereafter. We define the reduced thickness as it is proportional to the initial energy (or entropy) density of a collision at mid-rapidity, though the constant of proportionality is fixed by matching final multiplicities and is left to future work. The reduced thickness, $T_R$, is given by 
\begin{equation}
    T_R(\vec{r}) = \left(\frac{\tilde{T}_A(\vec{r})^{\,p} + \tilde{T}_B(\vec{r})^{\,p}}{2}\right)^{1/p}
\end{equation}
where the choice of $p$ determines the mathematical structure of the fluctuating density map. For the majority of this work, we assume $p=0$ as the default value as supported by prior Bayesian analyses~\cite{Bernhard:2019bmu,Giacalone:2021clp}.

\section{Sampling and Tuning}
\label{sec:sampling}

In this section, we outline the procedures we implement in order to construct the \ftrento model based on the physical framework described above. To sample parton momentum fractions and subsequently determine parton numbers in nucleons, we utilize the method of inverse cumulative distribution function (CDF) sampling. We begin by establishing the probability of a parton to carry longitudinal momentum fraction $x$ as
\begin{equation}
p_i(x,Q^2) = \frac{f_i^{\pA}(x,Q^2)}{N_i(Q^2)},
\qquad \int_{x_\mathrm{min}}^{1}p_i(x,Q^2)\,dx = 1, 
\label{eq:pi}
\end{equation}
where $N_i$ is the normalization factor defined by 
\begin{equation}
N_i(Q^2) = \int_{x_\mathrm{min}}^{1} f_i^{\pA}(x,Q^2)\,dx.
\label{eq:Ni}
\end{equation}
The value $N_i$ can also be considered the average number of partons of flavor $i$ per nucleon with momentum fraction $x > x_\mathrm{min}$.

The probability density $p_i(x,Q^2)$ in Eq.~\eqref{eq:pi} defines the cumulative distribution function
\begin{equation}
F_i(x,Q^2) = \int_{x_\mathrm{min}}^{x} p_i(x',Q^2)\,dx',
\label{eq:Fi}
\end{equation}
such that the momentum fraction of a single parton is sampled as $x = F_i^{-1}(\eta)$, where $\eta \sim U(0,1)$. We then sample the momentum fraction of valence quarks using the bound-proton nPDFs for the up ($u$) and down ($d$) valence quarks defined by
\begin{equation}
\begin{split}
    u_v^{\pA}(x) = f_u^{\pA}(x,Q^2)-f_{\bar u}^{\pA}(x,Q^2),\\
    d_v^{\pA}(x) = f_d^{\pA}(x,Q^2)-f_{\bar d}^{\pA}(x,Q^2),
    \label{eq:valences}
\end{split}
\end{equation}
respectively, with $u^{\pA}_v$ denoting the up quark valence distribution function, $d^{\pA}_v$ being the down quark valence distribution function, and subscript $u, d, \bar u, \bar d$ corresponding to the distributions of up, down, anti-up, and anti-down quarks, respectively. The momentum fraction of the valence quarks in a proton are 
\begin{equation}
\{x_1,x_2,x_3\}
= \big\{F_{u_v}^{-1}(\eta_1),\,F_{u_v}^{-1}(\eta_2),\,F_{d_v}^{-1}(\eta_3)\big\},
\end{equation}
where $\eta_1, \eta_2, \eta_3 \sim U(0,1)$ are independent uniform draws. We then build up the virtual sea one constituent at a time. We first draw a species from a categorical draw in $\mathcal{S}=\{g,u\bar u,d\bar d,s \bar s,c \bar c,\dots\}$, and then sample one momentum fraction for a gluon or two for a quark--antiquark pair using the corresponding CDF. The categorical draw from this species list requires that we define 
\begin{equation}
\lambda_i = \frac{N_i}{\sum_{j\in\mathcal{S}}N_j},
\end{equation}
in which $\lambda_i$ represents the fraction of a nucleon's sea constituents carrying flavor $i$.

\begin{figure}[!tbp]
    \centering
    \includegraphics[width=1\linewidth]{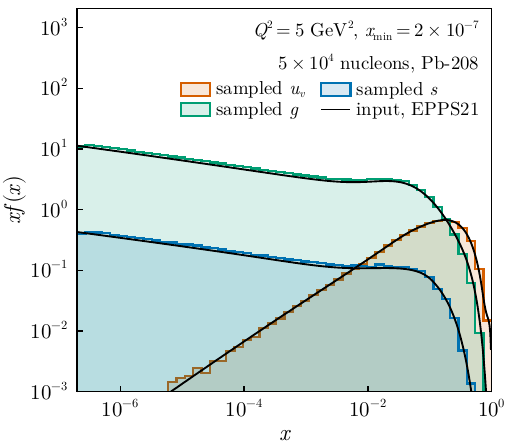}
    \caption{Sampled distributions of valence up quarks ($u_v$) in orange, gluons ($g$) in green, and strange quarks ($s$) in blue in comparison to the input bound-proton nPDF momentum distributions from EPPS21 (black lines) for Pb-208 at $Q^2 = 5$~GeV$^2$. The sampled histograms follow the expected $xf(x)$ for all $x_\mathrm{min} < x < 1$, with a small excess at the lowest $x$.}
    \label{fig:SamplingClosure}
\end{figure}

We track the total momentum fraction carried by all partons via
\begin{equation}
x_\mathrm{tot}=x_1+x_2+x_3+\sum_{k\,\in\,\mathrm{sea}}x_k,
\end{equation}
and stop sampling new constituents once $x_\mathrm{tot} \gtrsim 1$, after which we renormalize the momentum fractions through $x_k \ \rightarrow\ \frac{x_k}{x_\mathrm{tot}}$, which ensures that $\sum_k x_k = 1$ as expected. In Fig.~\ref{fig:SamplingClosure}, we show that the nPDF and the sampled parton distributions from our model agree for the full range of $x$, and each sampled species carries its expected share of the total momentum fraction to within 1\%. At small $x$, there is a constant shift from the stopping rule and the renormalization, though this effect is negligible in our observables. 

\begin{figure}[!tbp]
    \centering
    \includegraphics[width=1\linewidth]{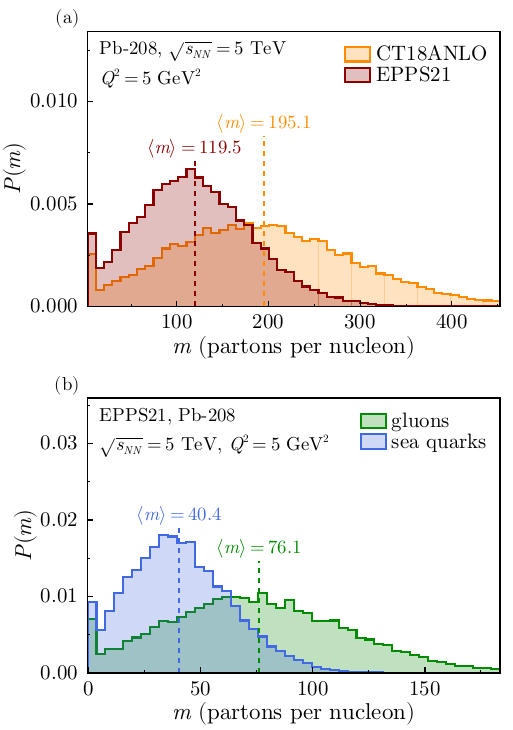}
    \caption{(a) Number of partons per nucleon comparing CT18ANLO~\cite{Hou:2019efy} PDFs with EPPS21~\cite{Eskola:2021nhw} nPDFs. (b) EPPS21 nPDF parton count further divided into counts of gluons and sea quarks. Both panels are calculated at $\sqrt{s_{NN}} = 5$~TeV and $Q^2 = 5$~GeV$^2$. Nucleons sampled from PDFs have 195.1 partons on average across $2\times 10^4$ sampled nucleons, while those from nPDFs have 119.5 partons on average. Of these 119.5, 76.1 are gluons and 40.4 are sea quarks and antiquarks (about 20 pairs). Panel (a) shows the effect of shadowing on parton counts while panel (b) depicts the dominance of gluons making up nucleons. For each plot, there is a small peak at 3 partons per nucleon, representing the nucleons in which all momentum is contained within the three valence quarks.}
    \label{fig:Parton_Hists}
\end{figure}

\begin{figure}[!tbp]
    \centering
    \includegraphics[width=1\linewidth]{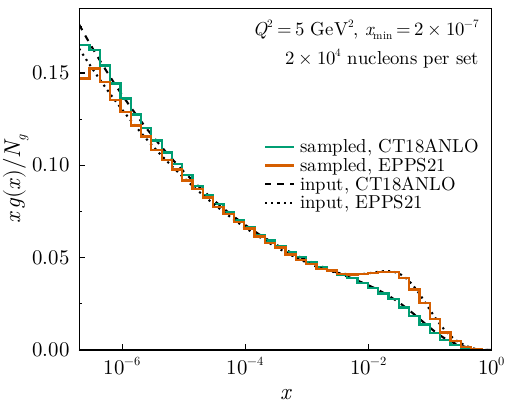}
    \caption{The sampled momentum-weighted gluon distribution reproduces the input gluon densities for both free-proton PDFs and nPDFs. It also shows the rising gluon number at low momentum fractions, and the significance of anti-shadowing on the gluon number for $x\gtrsim10^{-2}$ in nPDFs.}
    \label{fig:GluonCheck}
\end{figure}

The resulting distributions of partons per nucleon are shown in Fig.~\ref{fig:Parton_Hists}(a) for PDFs and nPDFs, with the nPDF parton count further divided into gluons and sea quarks in Fig.~\ref{fig:Parton_Hists}(b). These parton counts are calculated at $\sqrt{s_{NN}} = 5$~TeV and $Q^2 = 5$~GeV$^2$. We find that nucleons which have partons sampled from proton PDFs carry a significantly larger number of partons on average than those sampled from nPDFs. This is explained by the effect of shadowing for partons in heavier element nuclei~\cite{Eskola:2021nhw,Frankfurt:2011cs}. Furthermore, we find that gluons dominate the overall parton count in nPDF-sampled nucleons, because they are the most abundant partons at low momentum fractions, where most sampled partons lie. Additionally, we find the sampled gluon momentum fractions reproduce the input gluon densities for both free-proton PDFs and nPDFs, as shown in Fig.~\ref{fig:GluonCheck}. This plot further illustrates the buildup of gluon counts at lower momentum fractions, while simultaneously showing the effect of anti-shadowing on the gluon number for $x\gtrsim10^{-2}$ in nPDFs.

Thus far, all of the distribution functions used to sample partons have been from proton distributions in lead (Pb-208). In order to properly construct the full Pb-208 nucleus, we perform an isospin flip on $N=126$ of the nucleons, where the neutron number is $N = A - Z$ with $Z=82$ being the proton count and the mass number $A = 208$ for lead. As mentioned, we then distribute protons and neutrons via two distinct Woods--Saxon distributions. A simple rejection sampling procedure yields a list of $\vec r_n$ for each nucleus with length $A$, corresponding to the total number of nucleons in a nucleus.

From the Rayleigh distribution of partons around these nucleon centers, we define the transverse position of partons 
\begin{equation}
\vec{r}_k = \vec{r}_n + \delta_k
\begin{pmatrix} \cos\varphi_k \\ \sin\varphi_k \end{pmatrix}
\end{equation}
where $\delta_k$ is the radial offset parameter from the nucleon center. An inverse CDF of the Rayleigh distribution defines this value as
\begin{equation}
\delta_k = W(x_k)\sqrt{-2\ln(1-\eta)},
\end{equation}
for $\eta\sim U(0,1)$. Then, we perform a uniform draw of polar angles using $\varphi_k = 2\pi \eta'$ with $\eta'\sim U(0,1)$, defining the angular position of the parton. Now, the nucleon width for distributing partons $W(x_k)$ requires careful handling. 

Partons carrying different momentum fractions are distributed with different widths about the nucleon center, with high momentum valence quarks concentrated at the nucleon center, and low-$x$ partons spread to wider effective widths. One explanation of this phenomenon is the Gribov diffusion of low-$x$ partons in a virtual sea~\cite{Gribov:1973jg,Frankfurt:2005mc}. Partons in this regime undergo a random-walk-like diffusion in transverse space, with each step spreading the spatial distribution of the parton to a wider effective width. In this picture, low-$x$ partons can be considered the end of an emission cascade from a parent parton of larger $x$, such that the number of steps grows as $\ln(x_0/x)$ and the mean-square transverse spread of a parton grows linearly with it~\cite{Gribov:1973jg}. Following the work in Refs.~\cite{Frankfurt:2010ea,Strikman:2003gz,Strikman:2009bd,Schlichting:2014ipa,Hirvonen:2024zne}, we choose to model the width of the parton distribution about the nucleon center as a function of parton momentum fraction $x$
\begin{equation}
W(x) =
\begin{cases}
\sqrt{w_0^2 + 2\alpha'_g \ln\!\left(\dfrac{x_0}{x}\right)}, & x < x_0, \\[8pt]
w_0, & x \ge x_0,
\end{cases}
\label{eq:wx}
\end{equation}
with $w_0$ the nucleon width for non-diffusing partons, $\alpha'_g = 0.16$~GeV$^{-2}$ as the gluon Regge slope from HERA J/$\psi$ data~\cite{H1:2005dtp,ZEUS:2004yeh,Kowalski:2006hc,Rezaeian:2012ji,Kopeliovich:2007pq}, and $x_0 =0.01$ defining the momentum fraction at which we consider the onset of Gribov diffusion. We use the effective nucleon width $w=0.4$~fm~\cite{Giacalone:2021clp,Caldwell:2009ke,Schenke:2012wb,Nijs:2022rme}, and the parton size $\nu = 0.15$~fm~\cite{Mantysaari:2016jaz,Giacalone:2021clp,Hirvonen:2024zne,ALICE:2025luc,Kopeliovich:2007pq} here and in the following. Requiring that the momentum-weighted transverse width of a nucleon, including the parton width $\nu$, equals $w$ gives
\begin{equation}
w_0^2 = w^2 - \nu^2 - 2\alpha'_g \left\langle \max\!\left[\ln\!\left(\frac{x_0}{x}\right),\,0\right]\right\rangle_x,
\end{equation}
where $\langle\cdots\rangle_x$ denotes the average over the partons of $10^4$ sampled nucleons with each parton weighted by its momentum fraction $x_k$, evaluating to $w_0 = 0.37$~fm for $w=0.4$~fm and $\nu = 0.15$~fm.

\begin{figure*}[!tp]
    \centering
    \includegraphics[width=1\linewidth]{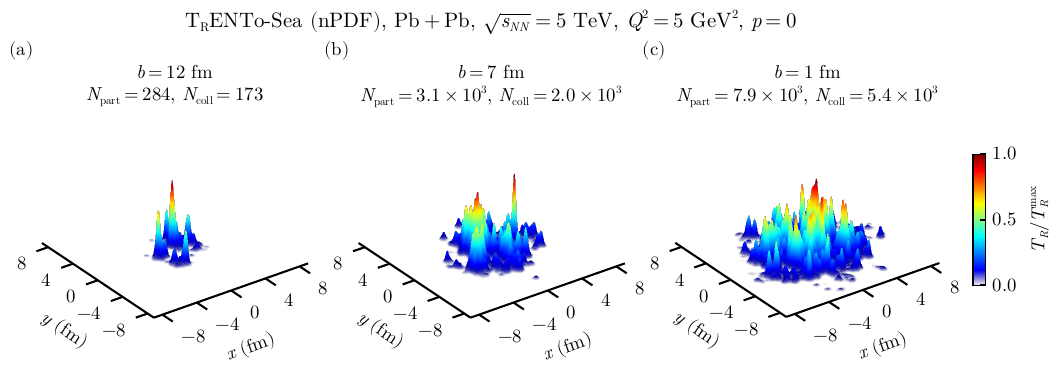}
    \caption{Different collision profiles for Pb+Pb collisions at $\sqrt{s_{NN}}=5$~TeV, $Q^2 =5$~GeV$^2$, and $p=0$. (a) Collision with impact parameter $b=12$~fm, representing a peripheral collision. (b) Collision with impact parameter $b=7$~fm. (c) The most central collision at $b=1$~fm, essentially direct overlap of colliding nuclei. The number of participant partons $N_\mathrm{part}$ grows with decreasing impact parameter, and the number of parton--parton collisions $N_\mathrm{coll}$ follows. In the central collision of (c), the two nuclei contain 52,270 partons, of which 7,910 participate in 5,427 collisions, while 397 nucleons are wounded. For more central collisions, we see more peaks in the fireball with a more crowded collision area.}
    \label{fig:Mountain_b_change}
\end{figure*}

To construct the initial conditions from participant partons we refer to Refs.~\cite{Miller:2007ri,Moreland:2014oya,Moreland:2019szz}. First, we sample the impact parameter $b$ as in \trento~\cite{Moreland:2014oya}. Figure~\ref{fig:Mountain_b_change} shows the resulting example profiles for different impact parameters. Larger impact parameters, representing peripheral collisions, have fewer colliding partons and a smaller area for energy deposition, whereas central collisions with $b\to0$ have a much more energetic profile and a larger collision area. The numbers of participant partons $N_\mathrm{part}$ and parton--parton collisions $N_\mathrm{coll}$ shown in Fig.~\ref{fig:Mountain_b_change} are at the parton level, so they are much larger than nucleon-level values. Unlike in nucleon-level models, $N_\mathrm{coll}$ is smaller than $N_\mathrm{part}$, because each collision involves two partons ($N_\mathrm{coll} = \langle n_\mathrm{hit}\rangle N_\mathrm{part}/2$), and we find $\langle n_\mathrm{hit}\rangle = 1.37$ collisions per participant in the central event of Fig.~\ref{fig:Mountain_b_change}(c). The parton with the most collisions in an event is always an energetic gluon, which typically collides with 7 other partons and with at most 10.

Colliding partons interact via a channel-specific effective cross section, given in Eq.~\eqref{eq:sigma}. We tune $s_0$ over $5\times10^4$ $pp$ simulated events, with partons sampled from free-proton PDFs, comparing the rate of collisions to total events as a probe of the inelastic cross section measured in experiments. In order to match the measured $\sigma_\mathrm{inel}^{pp}$~\cite{Loizides:2017ack,dEnterria:2020dwq}, we scan a parameter space of $s_0$ between 2 and 20~GeV$^2$, and a recursive bisection algorithm finds $s_0 = 6.28$~GeV$^2$ for $\xi = 2$ at $\sqrt{s_{NN}}=5$~TeV. We can then calculate $\sigma_\mathrm{inel}^{\mathrm{PbPb}}$ with partons sampled from nPDFs, and we find that $s_0 = 6.28$~GeV$^2$ yields an inelastic cross section for Pb+Pb collisions of $784 \pm 4$~fm$^2$, within the errors of ALICE's measured value of $767 \pm 25$~fm$^2$~\cite{ALICE:2022xir}, showing that our modeling of the effective parton-level cross sections is consistent with data. For values of $\xi <2$, we find that matching $pp$ inelastic cross sections requires a larger value of $s_0$ that also damps the cross sections at $\hat s \gg s_0$, where pQCD is reliable, eliminating collisions that we instead expect to occur. For $\xi = 1$, for example, the tuned cutoff is $s_0 \approx 29$~GeV$^2$. For $\xi\ge2$, the cross sections at $\hat s \gg s_0$ are left essentially unchanged, and our observables no longer depend on $\xi$. We therefore choose $\xi = 2$, the smallest value for which this holds. The remaining dependence of our observables on $\xi$ is tested in Sec.~\ref{sec:results}.

 We also find that the tuned value of $s_0$ is dependent on $\sqrt{s_{NN}}$. At $\sqrt{s_{NN}} = 200$~GeV, using our baseline $s_0$ value of 6.28~GeV$^2$ underestimates the Au+Au inelastic cross section by about 5\% in comparison to Glauber estimates of 680--685~fm$^2$~\cite{Loizides:2017ack,Giacalone:2019ldn}. Retuning $s_0$ to the $pp$ inelastic cross section of 4.2~fm$^2$ at $\sqrt{s_{NN}} = 200$~GeV~\cite{Loizides:2017ack,Giacalone:2019ldn} gives $s_0 = 3.19$~GeV$^2$ and a Au+Au cross section that agrees closely with these estimates. In contrast, using our fixed baseline $s_0$ to calculate the $pp$ inelastic cross section at $\sqrt{s_{NN}} = 200$~GeV underestimates it by close to a factor of two. Since nucleus--nucleus cross sections are largely determined by nuclear geometry, they are much less sensitive to $s_0$ than $pp$ cross sections. A change in $\sqrt{s_{NN}}$ therefore requires a recalibration of $s_0$, and we leave its energy dependence to future work.

\section{Results}
\label{sec:results}

\begin{figure*}[!tp]
    \centering
    \includegraphics{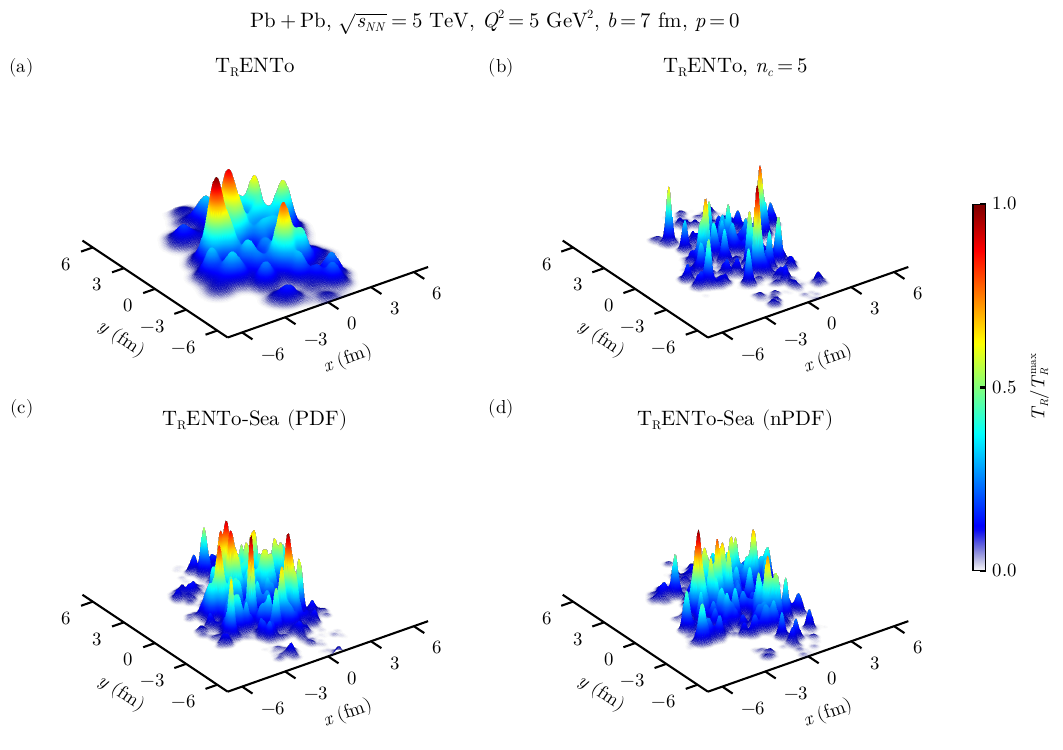}
    \caption{Comparison of initial-state models of colliding Pb-208 nuclei with $Q^2=5$~GeV$^2$ and $\sqrt{s_{NN}}=5$~TeV. For all models, we set nucleon width $w = 0.4$~fm and constituent/parton widths $\nu = 0.15$~fm for the collision at $b=7$~fm. (a) Nucleon-level \trento, with a smoother profile and a relatively small number of peaks. (b) Fixed-constituent \trento with $n_c=5$, where a few isolated hotspots dominate, with thinner peaks and more holes in the collision area. (c) \ptrento sampled from PDFs, where the number of peaks increases, more of them reach values close to $T_R^\mathrm{max}$, and the collision area is more filled in by collisions with low-$x$ partons. (d) \ftrento sampled from nPDFs, with more thin peaks than in (a) and (b), but fewer peaks close to $T_R^\mathrm{max}$ than in (c), which we attribute to the shadowing of low-$x$ partons in the bound proton nPDFs. The nucleon-level \trento model (a) uses gamma shape parameter $k = 1.19$, while the fixed-constituent \trento (b) has constituent number $n_c = 5$ with gamma shape parameter $k = 0.231$. The peaks in each plot are fractions between 0 and 1, since we show $T_R/T_R^\mathrm{max}$ where $T_R^\mathrm{max}$ is the maximum thickness within a single fireball.}
    \label{fig:Mountain_Models}
\end{figure*}

The aim of this paper is to develop a physically motivated initial-state profile of energy deposition using sub-nucleonic, fluctuating constituent numbers sampled from nPDFs. To understand the impact of our modeling choices on downstream observables, we compare the eccentricity harmonics $\varepsilon_n$, the normalized power spectrum $P(k)$, and the hole area fraction $\langle f_\mathrm{hole}\rangle$ of the simulated fireball with previous \trento models, and additionally compare the effects of sampling from free proton PDFs to nPDFs. A qualitative comparison of \ftrento to other iterations of \trento is given in Fig.~\ref{fig:Mountain_Models}. Here, we collide Pb-208 nuclei ($b=7$~fm) with $Q^2=5$~GeV$^2$ at beam energy given by $\sqrt{s_{NN}}=5$~TeV. The nucleon width $w$ is 0.4~fm for all models, and for each model containing sub-nucleonic structure we set $\nu = 0.15$~fm. Nucleons are given the same minimum distance condition at $d_\mathrm{min} = 0.4$~fm. The nucleon-level \trento model in Fig.~\ref{fig:Mountain_Models}(a) uses gamma shape parameter $k = 1.19$, corresponding to the maximum a posteriori fluctuation width $\sigma_\mathrm{fluct} = 1/\sqrt{k} = 0.918$ of Refs.~\cite{Bernhard:2018hnz,Bernhard:2019bmu}, while the fixed-constituent \trento in Fig.~\ref{fig:Mountain_Models}(b) has constituent number $n_c = 5$, within the posterior of Ref.~\cite{Moreland:2018gsh}, and fluctuation scale $\sigma_\mathrm{fluct} = 0.93$, the posterior median of Ref.~\cite{Moreland:2018gsh}, with gamma shape parameter $k = 1/(n_c\,\sigma_\mathrm{fluct}^2) = 0.231$. The peaks in each plot are fractions between 0 and 1, since we show $T_R/T_R^\mathrm{max}$ where $T_R^\mathrm{max}$ is the maximum thickness within a single fireball. These values are unitless and may be considered entropy or energy only after the introduction of the corresponding normalization factor.

\begin{figure}[!tbp]
    \centering
    \includegraphics[width=1\linewidth]{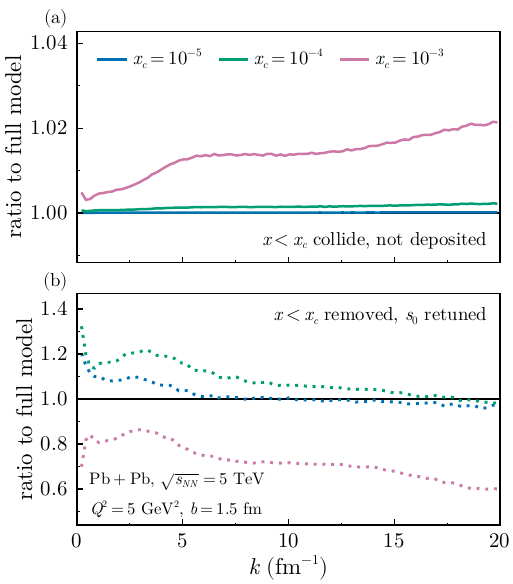}
    \caption{Ratio of the normalized power spectrum $\langle P(k)\rangle/\langle M\rangle^2$ to that of the full \ftrento (nPDF) model for Pb+Pb collisions at $\sqrt{s_{NN}} = 5$~TeV, $Q^2 = 5$~GeV$^2$, and $b = 1.5$~fm, averaged over $10^3$ events, for low-$x$ cutoffs $x_c$. (a) Partons with $x < x_c$ collide but do not deposit energy. (b) Partons with $x < x_c$ are removed before the collisions and $s_0$ is retuned to the $pp$ inelastic cross section.}
    \label{fig:LowXAblation}
\end{figure}

 First, introducing nucleon substructure leads to spikier energy density profiles. Each model uses the same nucleon geometry, so discrepancies between $T_R$ across models are due to parton-sampling and collision rate differences. The first initial-state \trento model does not have any structure below the nucleon width, and the energy deposition profile is smoother because of it. Upon the introduction of a fixed number of constituent particles, the number of peaks increases, and these spikes are noticeably sharper, since energy deposition occurs over a finer scale. Each constituent within a nucleon carries a gamma-distributed weight, and the small shape parameter $k = 0.231$ leads to large variations in the weight. A few constituents dominate the deposition to form hotspots that are more isolated than in other models. This model, as can be seen in the central overlap region, also has more holes in its profile than the nucleon-level model. For the \ptrento model sampled from PDFs, there is another increase in the number of energy deposition peaks, and more of these peaks reach values close to $T_R^\mathrm{max}$ than in the other models. The peaks also overlap more, leaving fewer regions without energy deposition. Nucleons sampled from PDFs have 195.1 partons on average, compared with 119.5 from nPDFs (Fig.~\ref{fig:Parton_Hists}), and the extra low-$x$ partons act as additional collision partners for higher-$x$ partons, whose energy deposition fills in the profile (see Fig.~\ref{fig:LowXAblation}). For the \ftrento sampled from nPDFs, the peaks overlap to a similar degree, but fewer of them reach values close to $T_R^\mathrm{max}$ than in the PDF model, which we attribute to the shadowing of low-$x$ partons in the bound proton nPDFs. The few dense overlap regions with a high-$x$ parton set $T_R^\mathrm{max}$, while the rest of the profile sits below it. In order to quantify some of these differences, we turn to calculations of the eccentricity.

\begin{figure*}[!tp]
    \centering
    \includegraphics[width=1\linewidth]{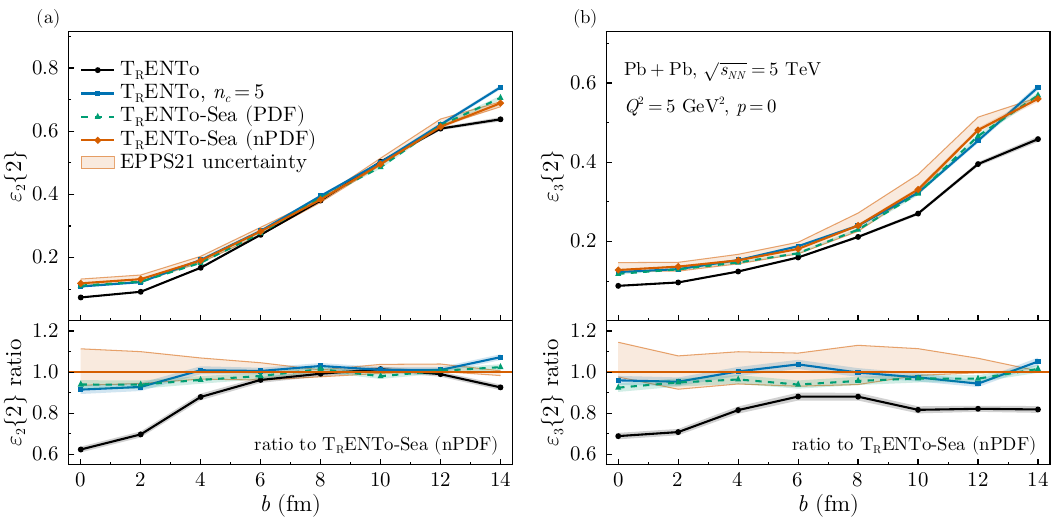}
    \caption{(a) $\varepsilon_2\{2\}$ as a function of $b$ for all models, averaged over $10^3$ collisions at each impact parameter, with its ratio to \ftrento with nPDFs shown below. The central collision differences are clearer in the ratio, where the nucleon-level model is 38\% smaller than our model and the fixed-constituent model about 9\% smaller, since in central collisions the eccentricity arises from fluctuations. (b) Same for $\varepsilon_3\{2\}$, with even better agreement among the sub-nucleonic models, but with nucleon-level \trento consistently lower in triangularity due to its lack of fine-structure fluctuations. The shaded orange band shows the EPPS21 nPDF theoretical uncertainty on \ftrento (nPDF). Statistical errors in this figure round to 0\% over the 10$^3$ collisions at each impact parameter.}
    \label{fig:EccModels}
\end{figure*}

We perform a Fourier decomposition of the repositioned fireball around its center of mass to find eccentricity harmonics. Following Ref.~\cite{Teaney:2010vd}, the eccentricity harmonics are
\begin{equation}
\varepsilon_n = \frac{\sqrt{\langle r^n \cos n\phi\rangle^2 + \langle r^n \sin n\phi\rangle^2}}{\langle r^n\rangle},
\label{eq:epsn}
\end{equation}
where $\langle\cdots\rangle$ denotes an average weighted by $T_R(\vec r)$, with $(r,\phi)$ measured from the centroid $\vec r_\mathrm{cm} = \langle \vec r\rangle$. For the model comparisons, $\varepsilon_n$ is computed event by event and averaged as an rms value over $N_\mathrm{ev}=10^3$ events per $b$. The rms value of the eccentricity harmonics is the spatial analogue to flow in momentum space, so we define
\begin{equation}
\varepsilon_n\{2\} = \sqrt{\langle \varepsilon_n^2 \rangle_\mathrm{ev}}
\label{eq:epsn2}
\end{equation}
as the parameter that maps to data~\cite{Giacalone:2020ymy}, where $\langle\cdots\rangle_\mathrm{ev}$ denotes an average over events rather than over the transverse plane. Its statistical uncertainty is $\delta\varepsilon_n\{2\} = \sigma(\varepsilon_n^2)/\big(2\sqrt{N_\mathrm{ev}}\,\varepsilon_n\{2\}\big)$, where $\sigma(\varepsilon_n^2)$ is the standard deviation of $\varepsilon_n^2$ over the $N_\mathrm{ev}$ events, and for ratios the uncertainties of both curves are added in quadrature. The shaded bands and error bars in our figures show these $\pm1\sigma$ statistical uncertainties.
In Fig.~\ref{fig:EccModels}, we show $\varepsilon_n\{2\}$ for $n = 2,3$ as a function of $b$ for all four models. Similar trends appear for higher order harmonics, but the eccentricities are not directly proportional to flow in this range due to nonlinear contributions.

The trends featured in Fig.~\ref{fig:EccModels} are noteworthy, in that we find all models agree on the ellipticity of a collision fireball within 5\% for $2<b<12$~fm. In the most central collisions, we find that there is a statistically significant difference between all four models, where \ptrento from PDFs and the fixed-constituent model are 6\% and 9\% different from \ftrento with nPDFs, and a nucleon-level model differs by around 38\% for these central collisions. The ratios in Fig.~\ref{fig:EccModels}(a) show this trend more clearly. For most impact parameters, the ellipticity $\varepsilon_2$ of a fireball is dominated by the collision geometry, namely the almond shape of two partially overlapping nuclei, whereas $\varepsilon_3$ is driven by fluctuations at all impact parameters. However, in central collisions, this geometry becomes a nearly circular fireball, for which the geometric contribution to $\varepsilon_2$ vanishes. This entails that in central collisions, nonzero eccentricities arise from fluctuations. These trends hold for higher order harmonics, but are statistically less significant, other than the separation of the nucleon-level model compared to sub-nucleon level models. The higher order harmonics, like that shown in Fig.~\ref{fig:EccModels}(b), are attributed to sub-nucleon fluctuations in general, so a model that does not have structure on this scale will automatically have lower eccentricities. In central collisions, we find that \ptrento with PDFs has a lower ellipticity than \ftrento from nPDFs, because there are a higher number of low-momentum fraction partons in the virtual sea. Without nuclear shadowing, there are more low-$x$ gluons in each nucleon, which serve as collision partners that allow more higher-$x$ partons to participate, and the energy deposited by these participants fills in geometric gaps in the fireball that would otherwise contribute to a higher eccentricity.

The shaded band in Fig.~\ref{fig:EccModels} shows the uncertainty propagated from the EPPS21 nPDFs. While the statistical uncertainty from the $10^3$ collisions at each impact parameter is nearly 0\%, the theoretical uncertainty is obtained by running our full collision pipeline at the upper and lower limits of the bound-proton nPDF errors. We use the same events as for the central set, so that the band reflects only the nPDF uncertainty. We find that the theoretical uncertainty is largest in central collisions, where $\varepsilon_2\{2\}$ changes by $-6\%$ to $+11\%$, and $\varepsilon_3\{2\}$ by $-2\%$ to $+15\%$. These uncertainties decrease toward peripheral collisions, and for $b \geq 8$~fm $\varepsilon_2\{2\}$ changes by less than 4\%. The band is largest in central collisions because the nPDFs are least constrained at small $x$, which sets the number of collision partners. The separation of the nucleon-level model from the sub-nucleonic models remains larger than the band, whereas the difference between \ftrento (PDF) and \ftrento (nPDF) lies within it. In central collisions, $\varepsilon_2\{2\}$ of fixed-constituent \trento sits below the band.

The above explanation does not suffice for a fixed-constituent model which is lower in ellipticity than \ftrento with nPDFs, because a fixed-constituent model has no partonic sea: every nucleon always has $n_c$ hotspots, and there are no soft gluons to fill in the collision profile. Instead, we find that there is an energy deposition correlation amongst hotspots in \ftrento that is not found in a fixed-constituent model. Collisions in \ftrento occur at the parton level alone, so dense overlap regions with a high momentum fraction parton tend to have more collisions. Across 50 events at $b = 1$~fm, the probability for a parton to collide rises from 0.5\% at $x \sim 10^{-7}$ to about 40\% at $x \sim 10^{-2}$, and participants with $x \gtrsim 10^{-2}$ collide on average about 1.5 times, compared with once for soft participants. The parton with the most collisions in each event is always a gluon, typically at $x \approx 0.05$, and collides 7 times in a typical event and at most 10 times. This is a source of event-by-event fluctuations, and is an alternative to the deposition of an entire nucleon which occurs in a fixed-constituent \trento model. If one nucleon is hit, all of its constituents deposit energy from their gamma weight regardless of position around the nucleon, smoothing the overall deposition profile compared to \ftrento.

\begin{figure}[!tbp]
    \centering
    \includegraphics[width=1\linewidth]{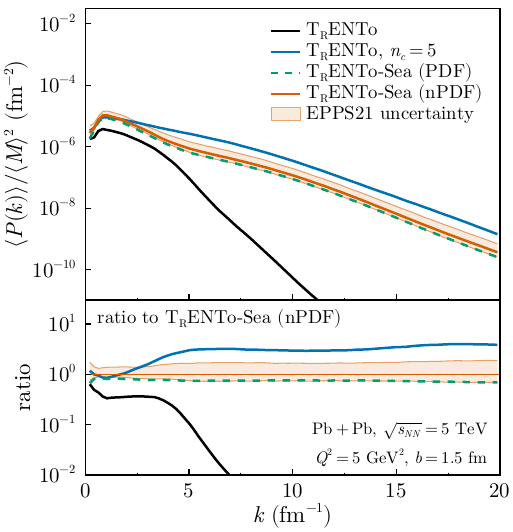}
    \caption{The normalized power spectrum ${\langle P(k)\rangle}/{\langle M\rangle^2}$ as a function of $k$, with its ratio to the nPDF \ftrento shown below. \ftrento models have more internal substructure (smaller lumps, higher $k$) than the nucleon-level \trento, which does not contain any structure smaller than a nucleon, and the power spectrum of the nPDF version is slightly larger than that of the PDF sampled \ptrento. The fixed-constituent \trento model has the most fine-scale structure, because it does not have any constituents other than its $n_c$ hotspots to fill in the space between them. The shaded orange band shows the EPPS21 nPDF uncertainty on \ftrento (nPDF).}
    \label{fig:PkModels}
\end{figure}

An analysis of the power spectrum allows a further investigation of sub-nucleonic structure across models, and the findings of Fig.~\ref{fig:PkModels} support the analysis of the findings in Fig.~\ref{fig:EccModels}. In order to calculate the power spectrum, we average over $10^3$ events, each recentered on its centroid, to get a mean profile $\bar T_R(\vec r) = \big\langle T_R(\vec r)\big\rangle_\mathrm{ev}$ which serves as a smooth background. Against this profile, every collision's variation from the mean has a unique decomposition,
\begin{equation}
\delta T_R(\vec r) = T_R(\vec r) - \bar T_R(\vec r)= \sum_{\vec k} \delta\tilde T_R(\vec k)\, e^{i\vec k\cdot\vec r},
\label{eq:deltaT}
\end{equation}
where $\vec k$ is the two-dimensional wave-number and $\delta \tilde T_R(\vec k)$ is the coefficient amplitude for the corresponding $k$. These amplitudes define the power spectrum over the transverse plane area $L_xL_y$,
\begin{equation}
\langle P(k)\rangle = L_x L_y \Big\langle \big|\delta\tilde T_R(\vec k)\big|^2 \Big\rangle_{|\vec k| = k,\ \mathrm{ev}},
\label{eq:Pk}
\end{equation}
where $\langle P(k)\rangle$ is the event averaged power spectrum. Here, the average is taken over events and over all wave vectors with the same magnitude $|\vec k| = k$. We normalize this value by the square of the event-averaged total deposition, $M = \int d^2r \, T_R(\vec r)$. We plot ${\langle P(k)\rangle}/{\langle M\rangle^2}$ for a range of $k$ in Fig.~\ref{fig:PkModels}, where large $k$ corresponds to small-scale structure.
By this interpretation, we find, as expected, that \ftrento models have more internal substructure (smaller lumps, higher $k$) than the nucleon-level \trento, and the power spectrum of the nPDF version is slightly larger than that of the PDF sampled \ptrento. However, we find that the fixed-constituent \trento model has the most fine-scale structure. Nucleon-level \trento has the least, since it does not contain any structure smaller than a nucleon. The EPPS21 nPDF uncertainty is propagated to the power spectrum in the same way as for the eccentricities. The band widens with $k$, from about $-14\%$ to $+42\%$ at $k < 2~\mathrm{fm}^{-1}$ to $-27\%$ to $+76\%$ at $k > 10~\mathrm{fm}^{-1}$. We find that the \ftrento (PDF) power spectrum sits at the lower edge of this band, and that the fixed-constituent model lies above the band for $k > 2.5~\mathrm{fm}^{-1}$. While not directly observable, the power spectrum has a measurable counterpart in the two-particle flow correlation $\langle v_n v_n\rangle$ across harmonics $n$, measured up to $n = 7$ in Pb+Pb collisions~\cite{ATLAS:2018ezv}. This is essentially a flow power spectrum, in which structure at larger $k$ appears in higher harmonics. However, at high harmonics viscosity damps the signal of fine-scale structure~\cite{Staig:2010pn,Staig:2011as} and nonlinear mixing between harmonics makes it difficult to quantify~\cite{Gardim:2011xv}, so only the small-$k$ part of the power spectrum is accessible in data.

\begin{figure}[!tbp]
    \centering
    \includegraphics[width=1\linewidth]{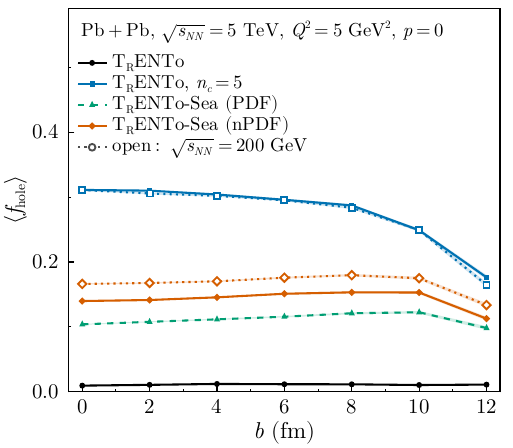}
    \caption{Hole area fraction $\langle f_\mathrm{hole}\rangle$, representing the share of the fireball area $\Omega$, defined in the text, in which energy deposition is nearly 0 as averaged over $10^3$ events. The nucleon-level \trento model has the lowest hole area fraction, close to zero across events, while a fixed-constituent model has the highest because it does not have a virtual parton sea. The \ftrento sampled from nPDFs has a slightly higher hole area fraction than the PDF sampled version. Open markers and dotted lines show the same models at $\sqrt{s_{NN}}=200$~GeV, where lowering the collision energy raises $\langle f_\mathrm{hole}\rangle$ for \ftrento but not for the fixed-constituent model.}
    \label{fig:Hole_Fraction}
\end{figure}

\begin{figure*}[!tp]
    \centering
    \includegraphics[width=\textwidth]{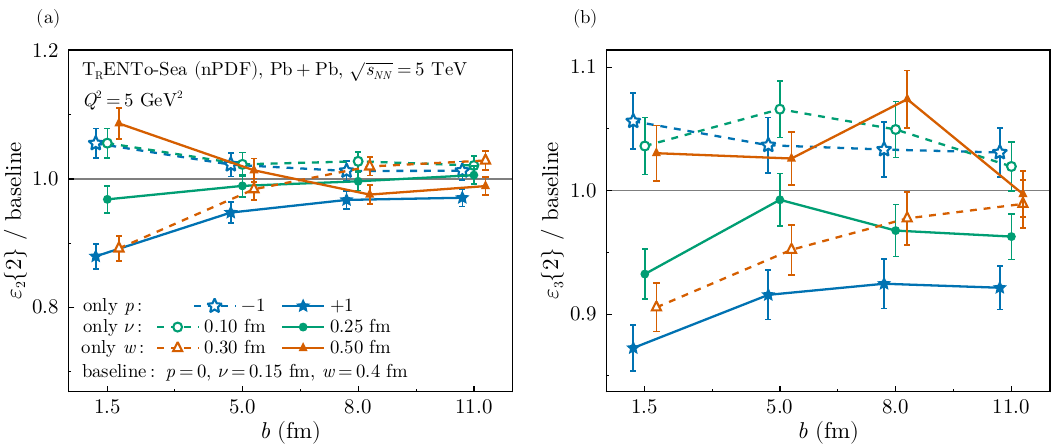}
    \caption{Ratio of (a) $\varepsilon_2\{2\}$ and (b) $\varepsilon_3\{2\}$ to baseline for a variety of different inputs. Each data point keeps two of the three parameters $p, w, \nu$ fixed at baseline ($Q^2 = 5$~GeV$^2$, $\sqrt{s_{NN}}=5$~TeV) while varying the free variable. At baseline, $p=0$, $w=0.4$~fm, and $\nu = 0.15$~fm. We test $p$ at both $-1$ and $+1$, $\nu$ at 0.1 and 0.25~fm, and $w$ at 0.3 and 0.5~fm, with the lower value shown by open markers and dashed lines and the higher value by filled markers and solid lines. The parameter $p$ is the only parameter that affects eccentricity across all impact parameters, while $w$ and $\nu$ change it only in central collisions, where varying $w$ has an effect comparable to $p$.}
    \label{fig:EccSensitivity}
\end{figure*}

\begin{figure*}[!tp]
    \centering
    \begin{minipage}[t]{0.49\textwidth}
        \centering
        \includegraphics[width=\linewidth]{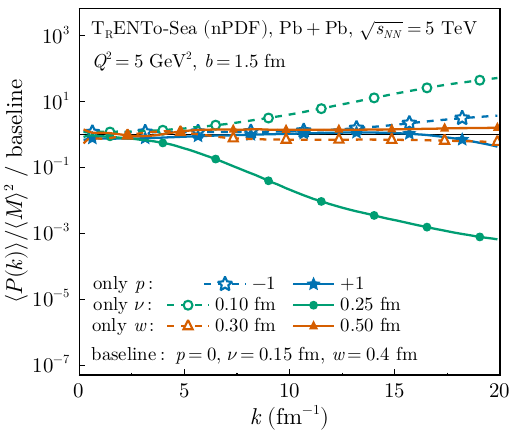}
        \caption{Ratio of the normalized power spectrum ${\langle P(k)\rangle}/{\langle M\rangle^2}$ for the $p$, $\nu$, and $w$ variants relative to baseline at $Q^2 = 5$~GeV$^2$ and $\sqrt{s_{NN}}=5$~TeV in Pb+Pb collisions. Setting $p = -1$ raises the power spectrum by at least 17\% at all $k$, while $p = +1$ lowers it. The parton width has the largest effect: $\nu = 0.10$~fm raises the power spectrum at large $k$ and $\nu = 0.25$~fm lowers it, by an amount that grows with $k$. Setting $w = 0.3$~fm lowers the power spectrum, while $w = 0.5$~fm raises it. Open markers and dashed lines show the lower value of each parameter, and filled markers and solid lines the higher value.}
        \label{fig:PkSensitivity}
    \end{minipage}\hfill
    \begin{minipage}[t]{0.49\textwidth}
        \centering
        \includegraphics[width=\linewidth]{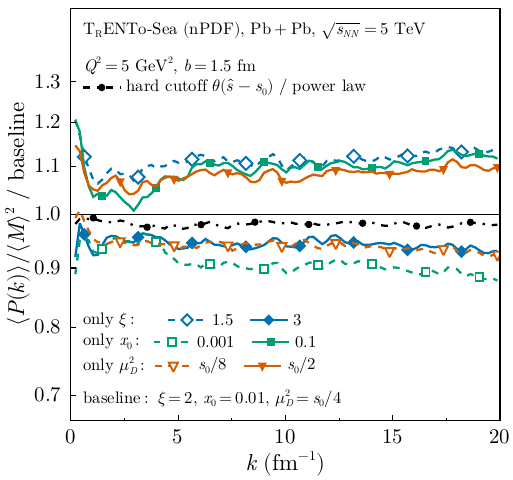}
        \caption{Ratio of the normalized power spectrum from different parameter variants for $\xi$, $\mu_D^2$, and $x_0$, compared to baseline with $\xi = 2$, $\mu_D^2 = s_0/4$, and $x_0 = 0.01$, with $s_0$ retuned to the $pp$ inelastic cross section for each variant. We test $\xi = 1.5, 3$, $\mu_D^2 = s_0/8, s_0/2$, and $x_0 = 0.001, 0.1$, with the lower value shown by open markers and dashed lines and the higher value by filled markers and solid lines. The black dash-dotted line with circles replaces the power-law damping factor with a hard cutoff $\theta(\hat s - s_0)$, where $\theta$ is the Heaviside step function, with $s_0 = 7.11$~GeV$^2$. Overall, the power spectrum changes by at most 12\%, but this change is nearly uniform across $k$, so the spectrum's shape is unchanged even though the total magnitude is affected by changes in these infrared parameters. Because $s_0$ is retuned for each variant, most of the effect of these parameters is absorbed into its calibration, leaving $s_0$ as the one tuned parameter of \ftrento.}
        \label{fig:IRSensitivity}
    \end{minipage}
\end{figure*}

The difference between \ptrento with PDFs and the fixed-constituent \trento when compared to nPDF \ftrento can be better understood through another quantity, namely the hole area fraction $\langle f_\mathrm{hole}\rangle$. This quantity, again determined over $10^3$ events, is defined by
\begin{equation}
    f_\mathrm{hole} = \frac{\mathcal{A}\big(\{\vec r \in \Omega : T_R(\vec r) < 0.02\,T_R^\mathrm{max}\}\big)}{\mathcal{A}(\Omega)}
\end{equation}
where $\Omega$ represents the fireball region in the transverse plane. We evaluate $T_R$ on a grid in the transverse plane with a spacing of 0.03~fm. For each $y$, $\Omega$ is accumulated between the first and last horizontal grid points where $T_R > 0.05\,T_R^\mathrm{max}$, so that $\Omega$ covers the fireball together with any gaps inside it. $\mathcal{A}(\cdot)$ is then the area of a region, given by its number of grid points multiplied by $(0.03~\mathrm{fm})^2$, and $f_\mathrm{hole}$ is the fraction of $\mathcal{A}(\Omega)$ where $T_R < 0.02\,T_R^\mathrm{max}$. In essence, this quantity determines the share of the fireball area in which energy deposition is nearly 0. As shown in Fig.~\ref{fig:Hole_Fraction}, we find that the nucleon-level \trento model has the lowest hole area fraction, close to zero across events, while a fixed-constituent model has the highest. The \ftrento sampled from nPDFs has a slightly higher hole area fraction than the PDF version. These trends, in agreement with the results in Fig.~\ref{fig:PkModels}, can be explained as follows. Fixed-constituent \trento has a higher hole area fraction because it does not have any constituents other than $n_c$. Alternatively, \ftrento models have many low-$x$ partons which collide probabilistically with higher-$x$ partons, allowing the higher-$x$ partons to deposit their energy, and fill in the energy deposition area from collisions. A PDF sampled \ptrento has a slightly lower hole area fraction than the nPDF version because shadowing reduces the number of partons in this low momentum region. This explains why \ftrento models have a smaller power spectrum at large $k$ than the fixed-constituent one despite having the same constituent width $\nu$. The hole area fractions at $\sqrt{s_{NN}}=200$~GeV, shown with open markers in Fig.~\ref{fig:Hole_Fraction}, support this explanation. We find that the fixed-constituent model without low-$x$ constituents has an identical hole fraction $\langle f_\mathrm{hole}\rangle$ at both energies, but lowering the collision energy raises $\langle f_\mathrm{hole}\rangle$ at every impact parameter for \ftrento. Referring back to Fig.~\ref{fig:transverse}, lower energy nuclei carry fewer low-$x$ partons to act as collision partners, which raises the hole area fraction. Propagating the EPPS21 uncertainty to the hole area fraction in the same way as for the eccentricities and the power spectrum (not shown in Fig.~\ref{fig:Hole_Fraction}) gives a large effect, with a band spanning from $-21\%$ to $+51\%$ in central collisions.

To isolate the role of low-$x$ partons, we perform an ablation test in Fig.~\ref{fig:LowXAblation} at $b=1.5\ \mathrm{fm}$. We define a cutoff momentum fraction $x_c$ below which partons do not deposit energy in this test. For the first variant shown in Fig.~\ref{fig:LowXAblation}(a), partons which carry $x<x_c$ collide as usual but do not deposit any energy as they would in the baseline \ftrento model. We find that the normalized power spectrum changes by less than 2\% across all wave numbers for $x_c \le 10^{-3}$. In the second variant shown in Fig.~\ref{fig:LowXAblation}(b), partons with $x<x_c$ are taken out of the collision system entirely, such that they do not collide with higher-$x$ partons nor do they deposit any energy. We find that removing these low-$x$ partons altogether changes the power spectrum by up to 18\% for $x_c = 10^{-4}$ and by 20--33\% when $x_c = 10^{-3}$. The inclusion of low-$x$ partons has a significant effect on the power spectrum (and acts similarly for the other observables), not because of the energy deposition from all of the low-$x$ partons, but because these partons act as collision partners for high-$x$ partons whose energy deposition matters in the collision fireball.
To test the robustness of these results against our choices of the model parameters $w$, $\nu$, and $p$, we ran a number of sensitivity tests. As outlined previously in this work, we use $w = 0.4$~fm, $\nu =0.15$~fm, and $p=0$ to get our results. In order to understand the effect that these choices have on our conclusions, we calculate $\varepsilon_2\{2\}$ and $\varepsilon_3\{2\}$ for a variety of different inputs, as well as ${\langle P(k)\rangle}/{\langle M\rangle^2}$ against these changes. For each test, two of the three parameters are held fixed at the value used for \ftrento, while either $p$ is tested for $-1$ or $+1$, $\nu$ is tested over 0.1 and 0.25~fm, or $w$ is changed between 0.3 and 0.5~fm. Each variant in the sensitivity scan re-calculates $w_0$, and recalibrates $s_0$ to $\sigma_\mathrm{inel}^{pp} = 6.7$~fm$^2$. The Pb+Pb cross section of each variant is then a consistency check against ALICE ($767 \pm 25$~fm$^2$)~\cite{ALICE:2022xir}, and all variants agree with it within 0.5 to 1.1 standard deviations. The exception is the parameter $p$, which needs no recalibration because it only enters through the reduced thickness mean after collisions have taken place. The baseline variant uses $s_0 = 6.28$~GeV$^2$, and the resulting sensitivity scans are displayed in Figs.~\ref{fig:EccSensitivity} and \ref{fig:PkSensitivity}.

The results in Fig.~\ref{fig:EccSensitivity} show that the parameter $p$ is the only parameter that affects eccentricity across all impact parameters, while $w$ and $\nu$ change it only in central collisions. At these low impact parameters, varying $w$ changes $\varepsilon_2\{2\}$ by up to 11\%, which is comparable with the 12\% change from $p$, while $\nu$ changes the eccentricities by 5\% at most. Setting $p=+1$ lowers $\varepsilon_2\{2\}$ and $\varepsilon_3\{2\}$ by about 12\% at central impact parameters and between 3--6\% in more peripheral collisions. Setting $p=-1$ raises the eccentricities by 1--6\%. This trend is expected from the reduced thickness generalized mean definitions in Ref.~\cite{Moreland:2014oya}. For an arithmetic mean, thickness profiles of both nuclei are additive~\cite{Moreland:2014oya}, so any time the energy deposition is zero in one nucleus, there may still be energy deposited at that point from the other, resulting in smoother profiles which in turn lower the eccentricity. The harmonic and geometric means vanish whenever either nucleus has zero energy deposited at a point, so they have spikier profiles with higher eccentricities. As $p$ decreases, the mean follows the smaller of the two thickness profiles, such that the harmonic mean ($p=-1$) gives even spikier profiles than the geometric mean ($p=0$), raising eccentricities between 1--6\%.

The effect of $\nu$ on $\varepsilon_2\{2\}$ and $\varepsilon_3\{2\}$ is nearly negligible, with $\nu = 0.10$~fm raising eccentricity between 2--5\% and $\nu = 0.25$~fm lowering $\varepsilon_3\{2\}$ in central collisions by at most $\sim 4$\%. Grainier profiles have higher eccentricities, and larger partons smooth out these fluctuations on the scale of a single parton. The parton size sets the finest geometric structure in our model; however, it does not change the number of energy-depositing peaks nor their position in the transverse plane. These factors set the eccentricity through its $r^n$ weighting. Accordingly, the same variations that change the power spectrum at $k > 10$~fm$^{-1}$ by orders of magnitude change it at $k < 2$~fm$^{-1}$, the scale relevant to the eccentricity, by only $+19\%$ and $-13\%$.

Varying the nucleon width, $w$, has an effect only in central collisions, with $w=0.3$~fm lowering $\varepsilon_2\{2\}$ by 11\% and $\varepsilon_3\{2\}$ by 9\%. Conversely, $w =0.5$~fm raises $\varepsilon_2\{2\}$ by 9\% and $\varepsilon_3\{2\}$ by 3\%. This result may seem counter-intuitive at first, however, lower values of $w$ cause a recalibration of $s_0$ to lower values, with $w=0.3$~fm yielding $s_0 = 3.10$~GeV$^2$ and $w=0.5$~fm calibrating $s_0 = 9.25$~GeV$^2$. The recalibration of $s_0$ to lower values will include more low-momentum parton collisions, such that $N_\mathrm{coll}$ increases and produces a profile with less granularity and more uniformly distributed energy deposition.

Now, we turn to the effect of the same variations in $p$, $\nu$, and $w$ on the normalized power spectrum, shown in Fig.~\ref{fig:PkSensitivity}. Setting $p = -1$ raises the power spectrum by at least 17\% at all $k$, and more than doubles it for $k > 10$~fm$^{-1}$, while $p = +1$ lowers it by at most 26\%, mostly for $k < 2$~fm$^{-1}$. Once again, this effect is explained by the mathematical structure of the arithmetic mean ($p=+1$), in which areas that have zero energy deposition at baseline can have energy deposition from the participant partons of only one nucleus. These areas fill in the gaps between peaks, resulting in smoother profiles, which reduce the fluctuations of the profile relative to its total deposition, most strongly at large scales. Since $p$ acts only after collisions have taken place, this smoothing comes from the mean itself and not from additional partons.

As expected, the parton width has the largest effect. Above $k = 10$~fm$^{-1}$, $\nu = 0.10$~fm raises the power and $\nu = 0.25$~fm lowers it by an amount that grows with $k$, reaching several orders of magnitude for $\nu = 0.25$~fm at $k = 20$~fm$^{-1}$. Since the parton size sets the finest geometric structure, as discussed for the eccentricities, smaller partons give a larger power spectrum at large $k$.

Setting $w = 0.3$~fm lowers the power at all $k$, while $w = 0.5$~fm raises it, by at most 33\% and 46\%, respectively. Once again, a change in $w$ leads to a corresponding change in $s_0$. When $w$ is smaller, so is the infrared cutoff scale $s_0$ from calibration, and this increases the number of parton-level collisions in our model. That leads to a smoother energy deposition profile with less small-scale structure than the baseline, which has fewer collisions and more separation between parton-level events.

We also test the sensitivity of our results against the infrared parameters, which control the treatment of low-$x$ partons and low-energy parton collisions, $\xi$, $\mu_D^2$, and $x_0$. At baseline, we choose values of 2, $s_0/4$, and 0.01, respectively. Similar to the treatment of $p, \nu$ and $w$, we vary these values and recompute the normalized power spectrum and the eccentricities. Each time that one parameter is changed, we tune $s_0$ to the proton-proton cross section in measurements as described previously, and recalculate $w_0$ if necessary. Testing over $\xi = 1.5, 2, 3$, $\mu_D^2 = s_0/8,s_0/4,s_0/2$, and $x_0 = 0.001, 0.01, 0.1$, we find that the second and third eccentricity harmonics change negligibly, by at most 7\% and within $2\sigma$ of their statistical uncertainty from baseline. This shows that changes in all infrared parameters are absorbed by the tuning of $s_0$ to data. However, in Fig.~\ref{fig:IRSensitivity}, we show the ratio of the normalized power spectrum from different parameter variants to baseline, and find that there is an effect of these parameters on the amount of structure in our model. In particular, we find that the power spectrum changes by at most 12\%, but that this change is essentially uniform in $k$. Therefore, the spectrum's shape is unchanged even though the total magnitude is affected by changes in the infrared parameters. For the collision profile, this means that the infrared parameters change the total magnitude of the fluctuations, but not the size of the structures that make them up. Since the power spectrum scales with the square of these fluctuations, a 12\% change in power corresponds to at most a 6\% change in their magnitude relative to the total deposition, so the profiles are only slightly grainier or smoother than at baseline. We also replace the power-law damping factor in Eq.~\eqref{eq:sigma} with a hard cutoff $\theta(\hat s_{kl}-s_0)$, where $\theta$ is the Heaviside step function. Retuning $s_0$ for this hard cutoff infrared regulator yields a value of $s_0 = 7.11$~GeV$^2$ instead of 6.28~GeV$^2$. This treatment has at most a 2\% change in the power spectrum compared to our power-law regulator at baseline, and the eccentricities change even less. Overall, these sensitivity tests show that for $\xi \ge 2$ our results are nearly insensitive to the choice of $\xi$, and that the other infrared choices are largely absorbed into the calibration of $s_0$ to data. Our model remains slightly sensitive to the scale at which pQCD stops working, which is set by $s_0$, the one parameter we tune. We also find that the observables are largely insensitive to changes in $Q^2$ for $Q^2 \geq 5$~GeV$^2$, though parton counts per nucleon do change as a consequence of nPDF sensitivity to $Q^2$.

\section{Conclusions and Outlook} 
\label{sec:conclusions}

In this work, we have developed a parton-level initial-state model for heavy-ion collisions called \ftrento, restricted to the initial-state geometry at mid-rapidity. Sampling from parton distribution functions in nuclei, we are able to distribute a fluctuating number of constituent partons within nuclei such that they deposit energy proportional to their longitudinal momentum fraction $x$. We follow the reduced thickness ansatz of prior \trento models, and find that sourcing event-by-event fluctuations from nPDFs rather than Bayesian-calibrated variables yields nearly the same eccentricities except for the most central collisions. We introduce two phenomenological expressions, $W(x)$ for nucleon width, and $(\frac{\hat s}{\hat s + s_0})^\xi$ to suppress low-momentum parton-level cross sections, where our one free parameter $s_0$ is tuned to proton-proton cross section data. We find that our modeling choices yield an internally consistent treatment of Pb+Pb cross sections with the same $s_0$ at a given collision energy, while a change in $\sqrt{s_{NN}}$ requires $s_0$ to be retuned. The sampled parton distributions reproduce the input nPDFs, and propagating the EPPS21 nPDF uncertainty through the full model gives an uncertainty on the eccentricities that is largest in central collisions. Testing the sensitivity of this model to changes in the parameter space shows that, within the ranges tested, the eccentricities are largely unaffected by parameter choices for $p=0, -1$ and by $\nu$, and change with $w$ only in central collisions, but change as expected for $p=+1$, following the same trend of eccentricity decreasing with increasing $p$ as in the original \trento~\cite{Moreland:2014oya}. Moreover, the power spectrum is sensitive to the size of parton constituents in the model, though this has negligible effect on the eccentricities. Further, we find that changes in the infrared parameters $\xi$, $\mu_D^2$, and $x_0$ can affect the overall magnitude of the normalized power spectrum but not its shape, and that for $\xi \ge 2$ these choices are largely absorbed in the calibration of $s_0$, leaving only a slight sensitivity to the scale at which pQCD stops working. Removing low-$x$ partons from the collisions shows that they shape the fireball as collision partners for higher-$x$ partons rather than through their own energy deposition. Lastly, we find that there are both qualitative and quantitative differences between fluctuations sourced by free proton PDFs and nuclear-modified nPDFs, although for the eccentricities this difference lies within the nPDF uncertainty. 

In the future, we plan to extend this model to 3D, preserving longitudinal information about these initial-state profiles for further investigation. Additionally, this work is in principle useful for the study of soft-hard correlations, where our model now includes the natural placement of jet production points inside a fluctuating fireball. Without the addition of new parameters, it is possible to correlate hard probes with the soft geometry surrounding them~\cite{Noronha-Hostler:2016eow}. A full validation of this model, in which the initial conditions are evolved with hydrodynamics and compared with measured flow coefficients $v_n$, is reserved for future work. These tests may also reveal potential constraints on nPDFs through the testing of different nPDF datasets in full collision pipelines.

\onecolumngrid
\begin{acknowledgments}

The authors want to thank Berndt Müller, Steffen A. Bass and Petja Paakkinen for useful discussions. This work is supported by the U.S. Department of Energy, Office of Science, Office of Nuclear Physics, grant No.~DE-FG02-05ER41367.

\end{acknowledgments}
\twocolumngrid

\bibliography{refs}

\end{document}